# Virtual stain transfer in histology via cascaded deep neural networks

Xilin Yang[1,2,3], Bijie Bai[1,2,3], Yijie Zhang[1,2,3], Yuzhu Li[1,2,3], Kevin de Haan[1,2,3], Tairan Liu[1,2,3], Aydogan Ozcan[1,2,3,4]

## Affiliations:

[1]Electrical and Computer Engineering Department, University of California, Los Angeles, CA, USA

[2]Bioengineering Department, University of California, Los Angeles, CA, USA

[3]California NanoSystems Institute (CNSI), University of California, Los Angeles, CA, USA

[4]Department of Surgery, David Geffen School of Medicine, University of California, Los Angeles, CA, USA

## Abstract

Pathological diagnosis relies on the visual inspection of histologically stained thin tissue specimens, where different types of stains are applied to bring contrast to and highlight various desired histological features. However, the destructive histochemical staining procedures are usually irreversible, making it very difficult to obtain multiple stains on the same tissue section. Here, we demonstrate a virtual stain transfer framework via a cascaded deep neural network (C-DNN) to digitally transform hematoxylin and eosin (H&E) stained tissue images into other types of histological stains. Unlike a single neural network structure which only takes one stain type as input to digitally output images of another stain type, C-DNN first uses virtual staining to transform autofluorescence microscopy images into H&E and then performs stain transfer from H&E to the domain of the other stain in a cascaded manner. This cascaded structure in the training phase allows the model to directly exploit histochemically stained image data on both H&E and the target special stain of interest. This advantage alleviates the challenge of paired data acquisition and improves the image quality and color accuracy of the virtual stain transfer from H&E to another stain. We validated the superior performance of this C-DNN approach using kidney needle core biopsy tissue sections and successfully transferred the H&E-stained tissue images into virtual PAS (periodic acid-Schiff) stain. This method provides high-quality virtual images of special stains using existing, histochemically stained slides and creates new opportunities in digital pathology by performing highly accurate stain-to-stain transformations.

## Introduction

Histochemical staining of tissue sections is a critical step in pathology, which serves as a gold standard for diagnosing various diseases[1]. For example, hematoxylin and eosin (H&E), the most commonly used histochemical stain[2], is relatively cost-effective and easy to access, which highlights the contrast between the cell nuclei and the extracellular matrix and helps provide important information about the pattern, structure, and type of cells. As another example, periodic acid-Schiff (PAS) stain highlights the glycated molecules and is commonly used to examine diseases of basement membranes[3]. However, traditional histochemical staining requires tedious tissue treatment steps and lab-based monitoring by a histotechnologist which can be laborious and costly, especially for special stains, which are often more challenging to perform compared to H&E. The tissue damage introduced during the chemical staining procedures is irreversible, thus making a wash-and-restain process very difficult. To overcome such limitations, *virtual staining* was developed using deep learning to generate computationally stained images from the images captured using *label-free* tissue slides. Such label-free staining methods have been demonstrated with autofluorescence imaging[4–6], hyperspectral imaging[7], quantitative phase imaging (QPI)[8], light-scattering imaging[9], reflectance confocal microscopy[10], and ultraviolet surface excitation microscopy[11], among others[12,13]. Virtual staining methods can reduce the time, labor and cost of tissue staining[14], enabling further molecular analyses to be performed on the same tissue slide as no destructive biochemical reaction is needed.

An alternative approach that can be used to overcome the disadvantages of traditional histochemical staining is virtually transforming the stain applied to a histochemically stained image into another stain (referred to as *stain transfer*), e.g., transferring H&E stain to PAS stain. Stain transfer introduces additional versatility to the concept of virtual staining. In clinical applications, after examining H&E-stained tissue sections, pathologists may suggest additional special stains to be acquired for a more accurate diagnosis. In such scenarios, when some types of stains are already prepared and imaged, and other types of stains are still required, virtual stain transfer can reduce the additional labor and cost of tissue preparation, staining and imaging. Deep learning-based stain transfer methods also possess the same advantages as virtual staining methods have, such as staining consistency and repeatability[15].

One of the main difficulties in training a deep neural network to perform the stain transfer lies in the data acquisition dilemma: a slide can only be stained once with one type of stain without a washing process which may damage the tissue, making the acquisition of paired images with different stains very challenging. Neighboring cut slides of the same tissue block do not provide pixel-wise matched image pairs. Therefore, the neighboring slides cannot be used in training supervised image translation models, such as pix2pix[16], due to the mismatch of the essential features or the inter-slide misalignment and distortions. The CycleGAN[17]-based approaches can potentially eliminate the requirement for pixel-wise paired data and can be used to obtain image translation models by learning latent consistency across unpaired image data. Several stain transfer methods using CycleGAN-like models with unpaired data have been demonstrated: from H&E to Masson's trichrome (MT)[18], to PAS and Jones' silver stain[19], from H&E to Ki-67 immunohistochemistry (IHC)[20], among others[21]. However, the lack of direct pixel-wise structural loss makes these CycleGAN-based models prone to potential hallucinations, thus deteriorating the output image quality[15,22]. To circumvent data acquisition difficulty, de Haan et al. [15] used pre-trained label-free virtual staining models to virtually transform the same autofluorescence image into two different stains concurrently, generating spatially-registered (i.e., perfectly-paired) image datasets. Using these perfectly paired datasets, the stain transfer models were trained to convert the existing histochemical stained images into other stain types. With the additional stain types provided by the stain transfer networks, the diagnosis accuracy of several non-neoplastic kidney diseases was improved[15]. However, these stain transfer models were trained using virtually generated images as the input and target, whose distributions inevitably deviate from the histochemically stained image data distribution. This is due to the limited data availability, network generalization, model inductive biases[23,24], and model uncertainty[25]. Such discrepancies might undermine the final performance of stain transfer networks.

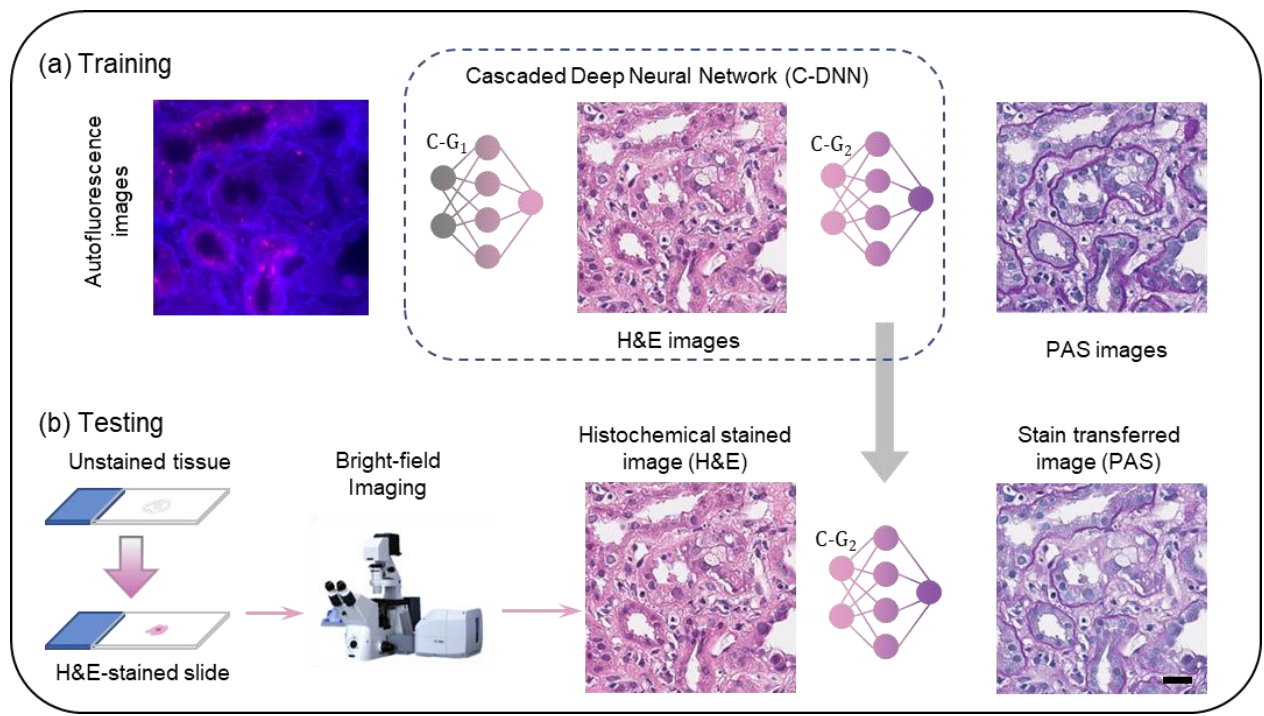

**Figure 1. Histochemical stain transfer via cascaded deep neural networks (C-DNNs).** The schematic illustrates the (a) training and (b) testing process of digitally transforming H&E-stained slide images into PAS-stained images. Autofluorescence (label-free), H&E and PAS-stained images are used in the training phase. Once the training is complete, only the second generator ($C\text{-}G_2$) is used for the blind testing, taking a histochemically stained H&E image as input and outputting a stain transferred PAS image as output. Scale bar: 30 μm

Here, we present a supervised deep learning-based stain transfer framework using a cascaded deep neural network, termed C-DNN (Figure 1), trained with two groups of data with different histochemical stains, as shown in Figure 2a. In the training process, two deep neural networks are cascaded to first virtually stain label-free autofluorescence images into H&E and then transfer the H&E stained output images into special stains. The cascaded network structure allows C-DNN to directly minimize the loss between the output images and the histochemically stained target images in *both* H&E and special stain domains, aiming to mitigate the discrepancy introduced by using input and target images generated by virtual staining networks and accordingly improve the stain transfer performance. After the training process of C-DNN, only the second neural network of the cascaded structure is used for blind stain transfer.

We demonstrated the success of C-DNN using kidney needle core biopsy tissue sections by transforming H&E-stained images into PAS stain. We compared our results with a single U-Net-based stain transfer model reported in Ref. [15] (referred to as the standard stain transfer network) both qualitatively and quantitatively, where our model achieved improved image quality, more accurate color distribution, and higher contrast. This method can provide high-quality virtual images of special stains from existing, histochemically stained slides, enriching the information available to pathologists with no additional cost of expert time and labor.

## Results
**C-DNN stain transfer framework and data preparation**

We demonstrated our framework by transferring H&E-stained images to PAS-stained images of the same field-of-view (FOV). To train the C-DNN framework, two groups of datasets are required, denoted here as group A and group B (see Figure 2). Each group contains three types of images: (1) autofluorescence images of the unstained/label-free tissue sections, (2) H&E-stained images (histochemically stained for group A and virtually stained for group B), and (3) PAS-stained images (virtually stained for group A and histochemically stained for group B). The virtually stained images in each group were generated from the label-free autofluorescence (AF) images using the corresponding pre-trained virtual staining networks, i.e., $G_{AF \to PAS}$ for group A and $G_{AF \to H\&E}$ for group B (see Figure. 2a and Methods section). These autofluorescence images in C-DNN training serve as a bridge to build perfectly-matched image pairs for both groups A-B, and they are no longer needed once the training process is completed; the model inference is a single feed-forward process of sub-second timescale per image tile.

The C-DNN contains two cascaded U-Net[26] structures (Figures 1 and 2b) as generators, each accompanied by a discriminator following the generative adversarial network (GAN)[27] scheme (see the Methods section). C-DNN training workflow is illustrated in Figure 2. In the training stage, the first generator takes autofluorescence images $x$ as input and outputs H&E-stained images $\widehat{y_1}$, denoted as $\widehat{y_1} = $ C-$G_1(x)$, where C-$G_1$ is the first generator of C-DNN, $y_1$ is the first target and $\widehat{y_1}$ is the first output. The second generator takes the output of the first generator and infers the PAS-stained output, denoted as $\widehat{y_2} = $ C-$G_2(\widehat{y_1}) = $ C-$G_2($C-$G_1(x))$. The first generator in the C-DNN structure converges to a label-free virtual staining network, while the second generator performs *virtual stain transfer*. At the testing stage, following the training, only C-$G_2$ is used by taking histochemical H&E-stained images as input to virtually infer PAS-stained images, denoted by $\widehat{y_2} = $ C-$G_2(y_1)$. To utilize the two groups of data (A and B, labeled by superscripts), the C-DNN is first trained on group A with equally weighted losses for C-$G_1$ and C-$G_2$ (see the Methods section). Upon convergence, we applied transfer learning to group B with adjusted weights. A higher weight is assigned for the loss from C-$G_2$ to emphasize the error between the histochemically stained PAS image and the second neural network output, where the error from C-$G_1$ is measured with virtually stained images. By using this cascaded structure and the two groups of data (A and B in Figure. 2), our framework has direct access during its training to the distances between its output images and the histochemically stained images for both H&E and PAS stains: $\mathcal{L}_1^A($C-$G_1(x^A), y_1^A)$ and $\mathcal{L}_2^B($C-$G_2($C-$G_1(x^A)), y_2^B)$ (see Figure 2b and the Methods section).

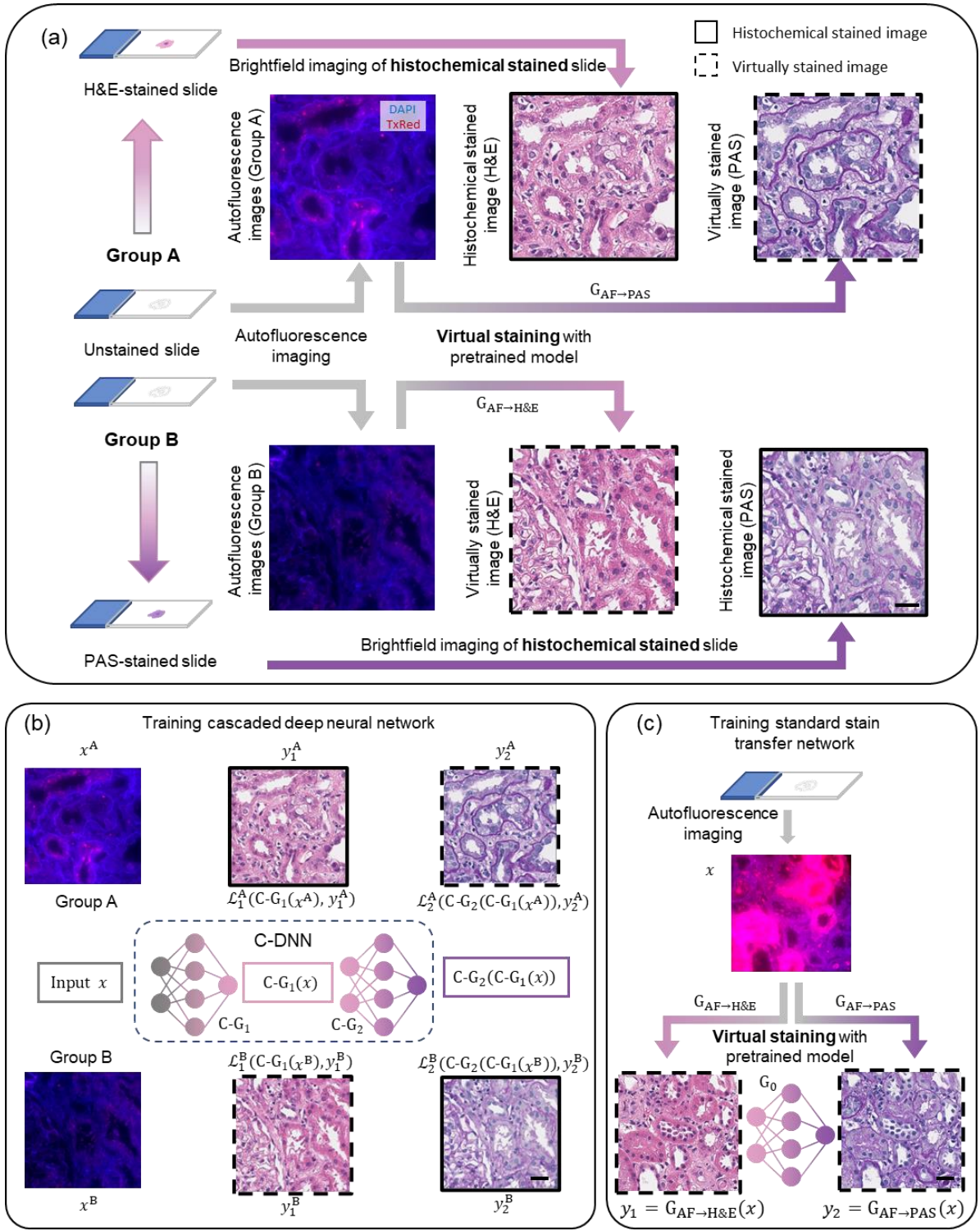

**Figure 2. Data preparation and training of C-DNN for virtual stain transfer.** (a) Data preparation: unstained slides are first scanned to obtain autofluorescence images (label-free) and then split into two groups, A and B. For group A, the unstained slides are histochemically stained with H&E, and the corresponding autofluorescence images are virtually stained to PAS using the pretrained virtual staining network $G_{AF \to PAS}$. For group B, the slides are histochemically stained with PAS, and the autofluorescence

images are virtually stained to H&E using the pretrained virtual staining network $G_{AF \rightarrow H\&E}$. (b) The C-DNN is trained on both group A and group B (labeled with superscripts) with the autofluorescence image used as the input $(x)$, H&E as the first target $(y_1)$ i.e., the virtual staining target, and PAS as the second target $(y_2)$, i.e., the stain transfer target. (c) The standard stain transfer relies on pretrained virtual staining networks to generate virtually stained images from autofluorescence images for both the input and output. Black solid boxes indicate histochemically stained images, while the dashed black boxes represent virtually stained images. Scalebar 30 μm.

**Stain transfer from H&E to PAS using C-DNN (C-$G_2$)**

We demonstrated our stain transfer method with C-DNN using two test datasets, from groups A and B. The testing FOVs were never seen by the networks in the training stage. The testing results and the comparison against the standard U-Net stain transfer (both the input and target are virtually stained) are summarized in Figures 3 and 4. In group A testing, histochemically stained H&E images were used as the input and the virtually stained PAS images were used as the target. Two testing FOVs are shown in Figure 3a. The C-DNN output images (C-$G_2(y_1^A)$) agree well with the PAS targets. C-DNN output images also reveal unique PAS features not available directly from the corresponding H&E images. For instance, in the virtually stained target of the first FOV (Figure 3, a1), we can observe basement membranes (labeled with yellow arrows) which are unique for PAS stain and are not clear in the input H&E images. The C-DNN output recovers these unique features, which agree well with the target, while the standard stain transfer network output ($G_0(y_1^A)$) could not recover such structures with high contrast. In the second FOV (Figure 3, a2), the standard stain transfer network stained the yellow-arrow-pointed region purple, which is inconsistent with the target, while the C-DNN output agrees well with the target color.

In group B testing, we used virtually stained H&E images as the input and the histochemically stained PAS images as the target. Two testing FOVs are shown in Figure 3b. The color of C-DNN output images is more accurate and provides a better match to the histochemically stained target, while the standard stain transfer network output color is denser and differs from the target. These color differences are further analyzed in Figure 4, supporting the same conclusion. To quantify the image structural similarity, we calculated the multiscale structural similarity index[28] (MS-SSIM, see the Methods section) between the output and target images for each FOV, the results of which are shown next to the output images in Figure 3. We also quantified the MS-SSIM over 20 different FOVs of $1024 \times 1024$ pixels for each group (from 6 patients for Group A, and 8 patients for group B), and the results are summarized in Table 1. This analysis indicates that the output images from C-DNN achieved higher MS-SSIM in both test groups.

|  | Group A | Group B |
| --- | --- | --- |
| C-DNN output, C-$G_2(\cdot)$ | 0.831 | 0.800 |
| Standard stain transfer, $G_0(\cdot)$ | 0.827 | 0.788 |

**Table 1.** MS-SSIM quantification of C-DNN compared with the standard stain transfer network.

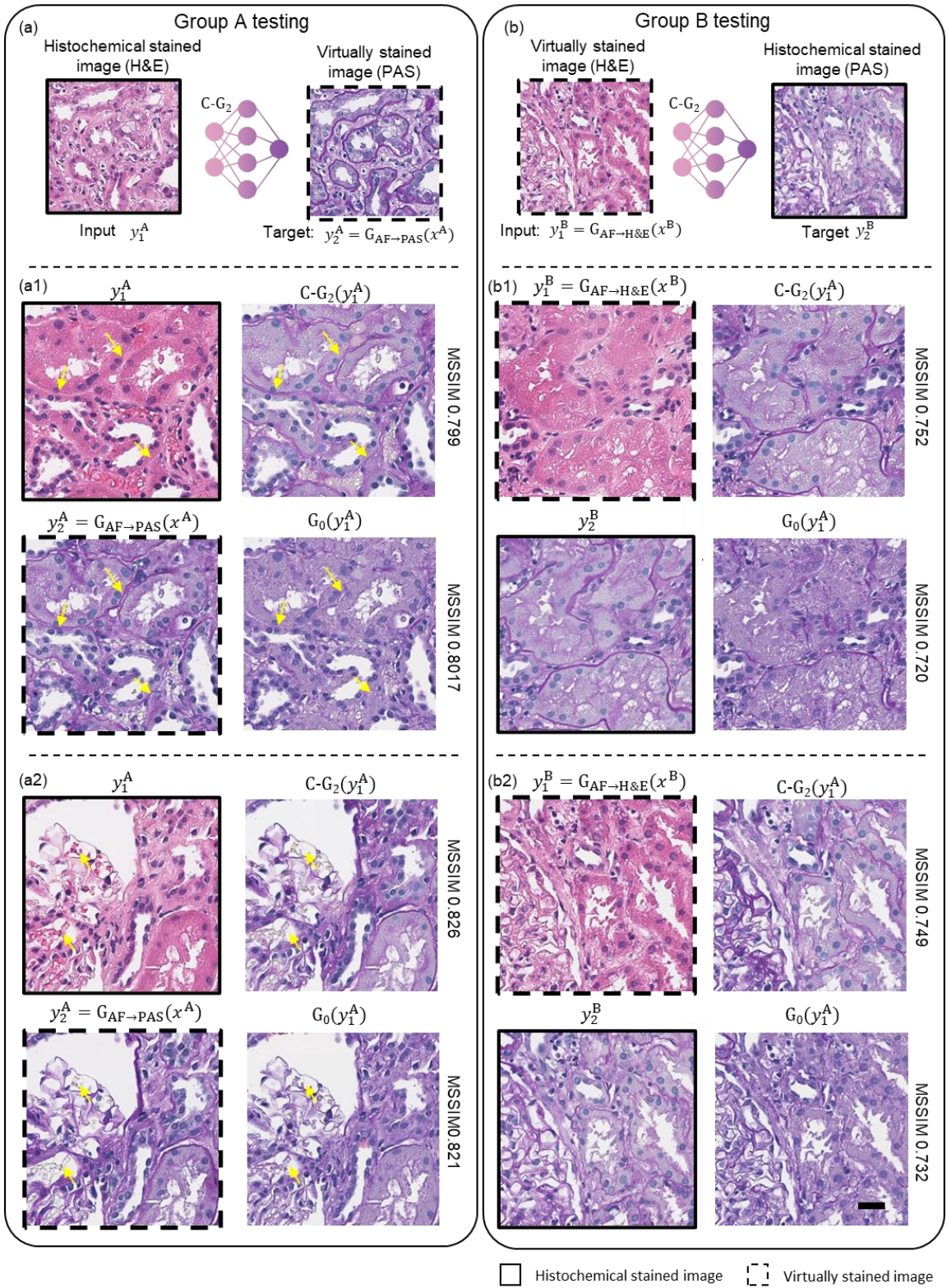

**Figure 3. Stain transfer from H&E to PAS using C-DNN and its comparison with the standard stain transfer network.** (a) Group A testing results and (b) group B testing results. Two FOVs for each group are tested, and the input, target and output images are shown; images from different FOVs are

separated by dashed horizontal lines. Here, $y_1$ is the H&E-stained input image and $y_2$ is the PAS-stained target. MS-SSIM values between the target and the output images from C-$G_2$ (C-DNN) and $G_0$ (standard stain transfer) are shown next to their output images. Scalebar: 30 μm.

Since chromatic distinction among different tissue constituents serves as one of the most significant features for pathologists to interpret tissue sections, we further quantified the color distribution of output images to assess the stain transfer quality. We compared the color distribution differences on two additional testing FOVs, one from each group, by converting the images from RGB to YCbCr color space. In Figure 4, we plot and compare the histograms of Cb and Cr channel values of the outputs from C-DNN (blue), the standard stain transfer network (red), and the target images (green). For both Cb and Cr channels, we observe that the C-DNN distributions agree well with the target distributions. The histograms are tighter for the standard stain transfer network output images and have higher peaks, signs of inferior image color contrast.

## Discussion

We presented a stain transfer framework using cascaded neural networks with superior performance compared to the standard stain transfer methods. C-DNN output contains high contrast features in the target domain, such as the basement membranes in PAS stains, which agree well with the histochemically stained targets with improved color accuracy. Without additional sample extraction or chemical treatment steps that are laborious, slow and costly, this framework rapidly reveals more features from the existing stained tissue sections, virtually enriching the information available to pathologists.

The success of this framework comes from using two dataset groups during the training and the cascaded neural network structure to better exploit the histochemically stained image data in the training phase. These enable the model to directly minimize the loss function calculated from the network output and histochemically stained target images for *both* H&E and PAS domains, i.e. $\mathcal{L}_1^A(\text{C-}G_1(x^A), y_1^A)$ and $\mathcal{L}_2^B(\text{C-}G_2(\text{C-}G_1(x^A)), y_2^B)$; also see the Methods section. Note that the loss terms from C-$G_1$ are only used to update the parameters in C-$G_1$ in each iteration. However, due to the cascaded nature of C-DNN, the output images of C-$G_1$ are directly fed into C-$G_2$ as input, forming a joint optimization scheme to avoid local minima which can be encountered if we were to only optimize C-$G_2$ separately. Converging into local minima can stagnate the training process and hinder the model from being further optimized. Updating the parameters of C-$G_1$ in a given training iteration will change the output distribution of C-$G_1$ and therefore benefit the training of C-$G_2$ in the subsequent iterations. Similarly, if only C-$G_2$ were to be optimized with Group B data, the stain transfer network would converge to a local minimum where it converts a virtually stained input distribution to a histochemically stained output distribution. When the histochemically stained images are used as the inputs for real-life applications, the discrepancy between the virtual and histochemical input distributions might deteriorate the stain transfer quality, which can be avoided by applying the C-DNN structure and joint optimization process outlined in this work. Furthermore, to better generalize to and handle lab-to-lab variations observed in the histochemical staining style, a pre-trained C-DNN can be fine-tuned using few-shot transfer learning[29] with adjusted weights using limited data similar to Group A, but with a unique histochemical stain style for H&E.

In summary, various stain-to-stain transformations can be learned and blindly performed using the C-DNN framework to provide unprecedented information on perfectly registered channels of different stains to pathologists. This cascaded neural network-based stain transfer method will open up new opportunities in digital pathology and histology.

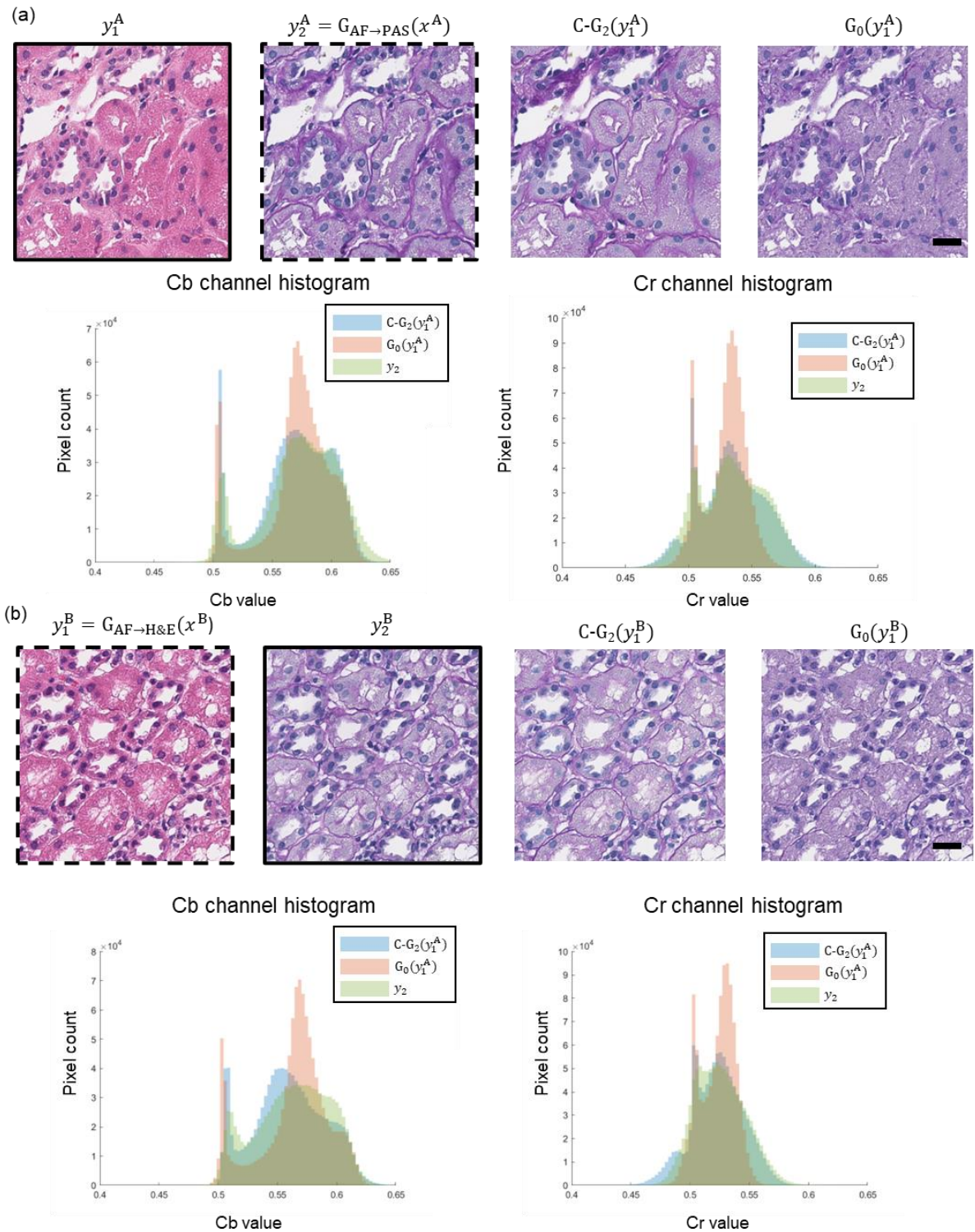

**Figure 4. Color distribution of stain transfer networks.** Two FOVs (one for each group, A and B) are tested, and the input, target and output images are shown for both groups. All the images are converted from RGB to YCbCr channels, and the Cb and Cr distributions are visualized with the histograms. (a) The testing images and their histograms of Cb and Cr values from group A, (b) testing images and their histograms of Cb and Cr values from group B. The histograms of the target images are plotted in green, and the histograms of C-DNN output images are plotted in blue, while for the standard stain transfer network, the corresponding image histograms are plotted in red. Scalebar: 30 μm.

## Methods

**Image acquisition and preprocessing**

The training data were acquired by microscopic imaging of thin tissue sections sliced from needle core kidney biopsies. The autofluorescence images were obtained using an Olympus IX-83 microscope with a DAPI and a Texas Red (TxRed) filter set (Semrock OSFI3-DAPI5060C, OSFI3-TXRED-4040C) using a 20x/0.75NA objective. The Tissue Technology Shared Resource performed the H&E and PAS histochemical staining at UC San Diego Moores Cancer Center. All the slides and digitized images were prepared from the existing specimen (under UCLA IRB 18-001029), and this work did not interfere with standard care practices or sample collection procedures. A bright-field scanning microscope (Leica Biosystems Aperio AT2 slide) was used to image the histochemically stained tissue sections with a 40x/0.75NA objective.

To create pairs of matched data for autofluorescence and bright-field images, we applied a multistage registration algorithm similar to previous works[10,15]. Bright-field images were first downsampled and coarsely matched to autofluorescence images using a correlation-based registration algorithm. Then a multi-modal registration algorithm was used to correct the rotation and image size of stained bright-field images with respect to the autofluorescence images. To address the local misalignment introduced in the histochemical staining process, we used a pre-trained virtual staining model to roughly stain the autofluorescence images. Then a pyramidal elastic registration algorithm[4,22,30] was implemented between the roughly stained autofluorescence and the stained bright-field images. The estimated transform was applied to the stained bright-field image to co-register it with respect to the autofluorescence image with sub-pixel accuracy.

After this co-registration process, the autofluorescence images were fed into two pre-trained virtual staining networks (described in the following sub-section) to acquire virtually stained images (PAS for group A, H&E for group B). For each group, we have three types of images for a training patch: autofluorescence image, virtually stained image, and histochemically stained image. These image patches are randomly cropped to a size of $256 \times 256$ pixels with ~10% overlap for training, and 40 image patches of $1024 \times 1024$ pixels are left for blind testing. The total number of the image patches used in the training phase is approximately ~25,000 for groups A and B together. In the training process, the dataset is split into training and validation sets with a ratio of 9 to 1.

**Implementation of the standard virtual staining networks**

The virtual staining networks used in the data preparation stage ($G_{AF \rightarrow H\&E}$, $G_{AF \rightarrow PAS}$) were a modified version of the original virtual staining network[4]. A progressive method was utilized in the training process to improve performance on edge details[31]. While generating virtually stained images, each image is zero-padded to $1536 \times 1536$ pixels to be compatible with the progressive GAN. An area of $1024 \times 1024$ pixels was cropped from the center FOV to avoid the edge effects induced by zero padding.

**Implementation of stain transfer deep neural networks**

The deep neural networks in this paper were trained using the GAN framework[27], each consisting of a generator (G) and a discriminator (D). The generator transforms the input $x$ into a target image $y$, while the discriminator network discriminates between the generator outputs and the target images to help guide the generator to produce images matching the distribution of the target data domain. The C-DNN structure contains two U-Net-based GANs cascaded such that the output of the first generator is directly used as the input for the second generator[32]. Both generators are composed of a five-block down-sampling path and a four-block up-sampling path. Down-sampling block contains two convolutional layers, each with kernels with a size of 3×3 and rectified linear unit (ReLU)[33] as activation function and residual connections[34], followed by a 2×2 average pooling layer with a stride of 2 to perform a down-sampling factor of 2. The up-sampling block is similar to the downsampling blocks, but concatenates the tensor of the last block with the tensor of the corresponding down-sampling path. The average pooling layers are replaced by 2× bilinear upsampling layers. The discriminator has six consecutive convolutional blocks with ReLU activations, followed by a fully-connected layer with a sigmoid activation function. All weights are initialized using Xavier[35].

The loss functions that C-DNN aims to minimize are:

$$\mathcal{L}_G = \mathcal{L}_{G,1} + \mathcal{L}_{G,2}$$
$$= -\log D_1\left(C\text{-}G_1(x)\right) + \alpha \cdot \text{MSE}(C\text{-}G_1(x), y_1) - \log D_2\left(C\text{-}G_2(C\text{-}G_1(x))\right) + \alpha \cdot \text{MSE}(C\text{-}G_2(C\text{-}G_1(x)), y_2) \quad (1)$$

$$\mathcal{L}_D = \mathcal{L}_{D,1} + \mathcal{L}_{D,2}$$
$$= -\log D_1(y_1) - \log[1 - D_1(C\text{-}G_1(x))] - \log D_2(y_2) - \log\left[1 - D_2\left(C\text{-}G_2(C\text{-}G_1(x))\right)\right] \quad (2)$$

where MSE refers to the 2D mean squared error, $C\text{-}G_i$ and $D_i$ represent the $i^{th}$ generator and discriminator, respectively, i.e., $i = 1\ or\ 2$.

For the standard stain transfer networks, the loss functions are:

$$\mathcal{L}_{G_0} = -\log D_0\left(G_0(y_1)\right) + \alpha \cdot \text{MSE}(G_0(y_1), y_2) \quad (3)$$

$$\mathcal{L}_{D_0} = -\log D_0(y_2) - \log[1 - D_0(G_0(y_1))] \quad (4)$$

Here, $y_1$ represents the H&E-stained image which serves as the input, $G_0$ and $D_0$ represent the generator and the discriminator, respectively. The networks are trained using an Adam optimizer[36] and $\alpha = 0.02$ with learning rates of $10^{-4}$ and $3 \times 10^{-5}$ for the generator and the discriminator networks, respectively.

C-DNN is first trained on group A data until convergence. Then we applied transfer learning to group B data with a weighted loss: $\mathcal{L}_G = \gamma \mathcal{L}_{G,1} + \mathcal{L}_{G,2}$, $\mathcal{L}_D = \gamma \mathcal{L}_{D,1} + \mathcal{L}_{D,2}$, where $\gamma = 0.1$. This weighted loss was used to help the C-DNN focus more on minimizing the loss between the virtually stained PAS output and the *histochemically* stained PAS images.

To ensure fair comparisons, the generators and the discriminators used in C-DNN and standard U-Net-based stain transfer networks are of the same structure and were trained using the same image data. The standard stain transfer network was trained with both groups A and B, using autofluorescence and virtually stained images. All the networks are implemented using TensorFlow[37] v2.7.0 and python version 3.9.7 with Compute Unified Device Architecture (CUDA) version 11.3.1. Training and testing are completed on a computer with two GeForce RTX 3090 graphics processing units (GPU) and an Intel Core i9-10920X central processing unit (CPU) with 256GB of random-access memory RAM. Typically,

~24 hours are needed for training convergence, and the blind inference time of $C$-$G_2$ for a $1024 \times 1024$ pixel image is ~0.65 seconds.

For performance evaluations, multiscale structural similarity index is used to quantify the quality of the network output images. The MS-SSIM[28] averages the structural similarity index (SSIM)[38] on multiple scales. The SSIM and MS-SSIM are calculated as:

$$\text{SSIM}(x,y) = \frac{(2\mu_x\mu_y + c_1)(2\sigma_{xy} + c_2)}{(\mu_x^2 + \mu_y^2 + c_1)(\sigma_x^2 + \sigma_y^2 + c_2)} \tag{5}$$

$$\text{MS-SSIM}(x,y) = \frac{1}{M}\sum_{i=1}^{M} \text{SSIM}(x_i, y_i) \tag{6}$$

where $\mu_x$ and $\mu_y$ are the mean values of the images $x$ and $y$, $\sigma_x^2$ and $\sigma_y^2$ are the variances of the images $x$ and $y$, respectively, and $\sigma_{xy}$ is the covariance between images $x$ and $y$. $c_1$ and $c_2$ were set to be $0.01^2$ and $0.03^2$. The subscripts of $x_i$ and $y_i$ represent downsampled images by $i$ times. MS-SSIM is averaged over eight scales (M=8) on images with $1024 \times 1024$ pixels.

To assess color distribution accuracy, the images were converted from RGB color space to YCbCr color space[39]. Each histogram in Figure 4 was collected from one image with $1024 \times 1024$ pixels.